\documentclass[conference]{IEEEtran}
\IEEEoverridecommandlockouts

\usepackage{cite}
\usepackage{amsmath,amssymb,amsfonts}
\usepackage{algorithmic}
\usepackage{graphicx}
\usepackage{textcomp}
\usepackage{xcolor}
\usepackage{svg}
\usepackage[caption=false,font=footnotesize]{subfig}
\makeatletter
 
\renewcommand{\p@subfigure}{\thefigure} 
 
\makeatother
\usepackage{booktabs}
\usepackage{multirow}
\usepackage{enumitem}
\usepackage[colorlinks=true, linkcolor=blue, citecolor=blue, urlcolor=blue]{hyperref}
\usepackage{aliascnt}

\def\BibTeX{{\rm B\kern-.05em{\sc i\kern-.025em b}\kern-.08em
    T\kern-.1667em\lower.7ex\hbox{E}\kern-.125emX}}

\begin{document}

\title{A SISA-based Machine Unlearning Framework for Power Transformer Inter-Turn Short-Circuit Fault Localization}
\author{
\IEEEauthorblockN{
    Nanhong Liu\IEEEauthorrefmark{1}, \textit{Graduate Student Member, IEEE,} 
    Jingyi Yan\IEEEauthorrefmark{1}, \textit{Graduate Student Member, IEEE,} \\
    Mucun Sun\IEEEauthorrefmark{2}, \textit{Member, IEEE,}
    and Jie Zhang\IEEEauthorrefmark{1}, \textit{Senior Member, IEEE}
}
\IEEEauthorblockA{
\IEEEauthorrefmark{1}The University of Texas at Dallas, Richardson, TX, USA\\
\IEEEauthorrefmark{2}Idaho National Laboratory, Idaho Falls, ID, USA \\
Email: {jiezhang@utdallas.edu} } }
\maketitle

\begin{abstract}
In practical data-driven applications on electrical equipment fault diagnosis, training data can be poisoned by sensor failures, which can severely degrade the performance of machine learning (ML) models. 
However, once the ML model has been trained, removing the influence of such harmful data is challenging, as full retraining is both computationally intensive and time-consuming. 
To address this challenge, this paper proposes a SISA (Sharded, Isolated, Sliced, and Aggregated)-based machine unlearning (MU) framework for power transformer inter-turn short-circuit fault (ITSCF) localization.  
The SISA method partitions the training data into shards and slices, ensuring that the influence of each data point is isolated within specific constituent models through independent training. 
When poisoned data are detected, only the affected shards are retrained, avoiding retraining the entire model from scratch. 
Experiments on simulated ITSCF conditions demonstrate that the proposed framework achieves almost identical diagnostic accuracy to full retraining, while reducing retraining time significantly. 

\end{abstract}


\section{Introduction}
In modern power and energy systems, data-driven machine learning (ML) models are playing a significant role in condition monitoring, fault diagnosis, and health assessment of critical electrical equipment such as generators and transformers \cite{yan2025bi, liu2025machine, zemouri2025power}. 
Generally, ML models rely heavily on high-quality datasets collected from sensors, which are assumed to be accurate and representative. 
However, in practical industrial environments, training data could be contaminated by sensor failures, which represent a pervasive and unavoidable reality in monitoring systems because of various degradation mechanisms, including electromagnetic interference (EMI), component aging, and complete sensor failures \cite{chandra2024sensor}.
% Communication systems may introduce data corruption through hardware aging, environmental conditions, calibration drift, electromagnetic interference (EMI), or software errors \cite{chauhan2024analysing}. 
% Additionally, labeling all training samples is expensive and prone to labeling errors, since human operators may introduce labeling errors during fault event annotation, particularly when distinguishing between similar fault signatures \cite{wang2023online}.

Before training, data preprocessing can be somewhat effective in removing clearly erroneous samples. 
However, in practical transformer monitoring systems, many data distortions are ambiguous.
Issues such as sensor failures often produce signal patterns that are difficult to distinguish from those caused by operating conditions or environmental influences during preprocessing \cite{belgacem2025toward}.
As a result, such contaminated samples cannot always be reliably identified before model training. 
When these data are used for training, ML models may suffer from degraded performance, leading to reduced accuracy. 
In real-world deployments, EMI-induced sensor failures are often identified only after model training, typically through system diagnostics, sensor consistency checks, or maintenance inspections.
At that stage, the influence of the poisoned data must be removed from an already trained model.
The most straightforward solution is to remove the affected samples and retrain the model from scratch. 
While effective, full retraining is computationally expensive and time-consuming. 
This limitation motivates the use of machine unlearning techniques that can eliminate the impact of poisoned data without completely retraining \cite{shaik2024exploring}.

% Before training, data preprocessing can be somewhat effective in removing clearly erroneous samples. 
% Nevertheless, in cases involving issues such as sensor drift or bias, EMI malfunction, and software errors, such anomalies are difficult to detect during preprocessing, especially when caused by sensor issues. \cite{belgacem2025toward}.
% When such poisoned data are used for training, the ML models suffer from degraded performance, resulting in reduced fault identification accuracy or even catastrophic misdiagnosis of severe faults.
% Once the ML model is trained, the most straightforward approach to address malicious or poisoned data is to remove the unwanted samples, followed by complete retraining from scratch. 
% Although full retraining is effective at removing all undesired data, this naive solution is prohibitively expensive in terms of both computational resources and time \cite{shaik2024exploring}.

To overcome these limitations, the concept of machine unlearning (MU) was first introduced to remove data traces by converting learning algorithms into a summation form \cite{cao2015towards}, and MU aims to efficiently remove the influence of specific training samples from a trained model without requiring full retraining.
Typically, MU algorithms are classified into two categories based on whether they fully remove the influence of unwanted data: exact unlearning and approximate unlearning \cite{li2025mubox}.
Approaches to approximate MU include influence function-based methods \cite{xu2024task}, which approximate the effect of removing data via inverse-Hessian computations, and gradient-based techniques that perform targeted parameter updates \cite{thudi2022unrolling}. 
However, these methods involve expensive matrix inversions or require careful tuning of hyperparameters, and are often less stable when applied to deep sequence models, limiting their adaptivity to certain MU scenarios.

Among exact MU algorithms, the SISA (Sharded, Isolated, Sliced, and Aggregated) methodology is a model-agnostic approach proposed by Bourtoule et al. \cite{bourtoule2021machine} that offers a particularly elegant solution by redesigning the training process itself to facilitate efficient unlearning. 
The key insight of SISA is to partition the training dataset into multiple independent shards, each of which is further divided into sequential slices. 
For each shard, constituent models are trained incrementally on larger subsets of data, and their predictions are aggregated to form the final ensemble output. 
Crucially, this architecture ensures that the influence of any individual data point is localized to a specific shard and slice, rather than being diffused across the entire model. 
When contaminated or poisoned data are identified, only the affected shard needs to be partially retrained, starting from the compromised slice, and efficiently reducing computational cost compared to full retraining.

In this study, we present a SISA-based MU framework for ITSCF localization in power transformers considering sensor failures. 
Our contributions are summarized as follows:
\begin{itemize}
    \item Propose a SISA-based MU framework using a softmax-probability averaging
    strategy.
    \item Develop a simulated ITSCF dataset with EMI sensor failures to evaluate accuracy and computational efficiency.
    \item Demonstrate that SISA unlearning restores diagnostic accuracy while reducing retraining time markedly.
\end{itemize}

The remainder of this paper is organized as follows.
\autoref{sec:simulation} describes the ITSCF simulation and dataset generation, including sensor failure modeling.
\autoref{sec:sisa_fw} introduces the proposed SISA-based MU framework for transformer ITSCF localization.
\autoref{sec:reaults} presents the experimental setup, performance evaluation, and analysis of the results.
Lastly, \autoref{sec:conclusion} concludes the paper.

\section{ITSCF Data Acquisition} \label{sec:simulation}
\subsection{ITSCF Simulation}
To obtain training data, a variety of power transformer ITSCF conditions under different severities were simulated using a 1.5 MW wind turbine model in MATLAB/Simulink. 
The simulation environment incorporates realistic wind speed profiles, generator dynamics, power electronics, and a three-phase generator step-up (GSU) power transformer with detailed winding models.

ITSCFs were systematically generated on both the low-voltage (LV) and high-voltage (HV) sides across all three phases A, B, and C, yielding a total of 48 distinct ITSCF conditions, with 8 conditions on both LV and HV sides in each phase.
Two-voltage-side three-phase current signals from the GSU transformer in power generation mode, lasting for a total of 15 seconds, were recorded as input data for the ML model.

\subsection{Sensor Failures}
In practice, current transformers (CTs) used for current measurement are susceptible to a variety of errors and failure mechanisms that can significantly compromise the reliability of the acquired signals. 
Among these factors, EMI is one of the most prevalent sources of sensor failures \cite{huang2024electromagnetic}.
Such interference can introduce high-frequency noise, impulsive spikes, amplitude bias, phase deviations, or even harmonics, all of which degrade the accuracy and fidelity of the measured current waveforms.

Consequently, EMI is introduced into the CT measurements on selected power transformer ITSCF conditions to simulate sensor failures, resulting in poisoned datasets.

\section{SISA Framework} \label{sec:sisa_fw}
\subsection{ML and MU Algorithms}
In this study, Long Short-Term Memory (LSTM) is adopted as the baseline ML model for ITSCF localization. 
Transformer current signals exhibit strong temporal dependencies, and fault-related signatures, such as phase current distortion, harmonic variations, and transient asymmetry, often evolve over short time windows. 
LSTM models are particularly suitable for this type of sequential data because they capture both short-term and long-term temporal patterns while avoiding the vanishing gradient problem that affects traditional recurrent neural networks (RNNs).

To efficiently remove the influence of EMI-poisoned samples from the training dataset, the SISA algorithm is employed. 
In SISA, the training dataset is divided into multiple independent shards, and each shard is further partitioned into ordered slices. 
Separate models are trained for each shard, and the outputs of all sharded models are aggregated to produce the final prediction.

This SISA architecture provides a major advantage for MU: when certain data needs to be removed, only the shards that contain those specific samples are retrained from scratch. 
All other trained shards remain unchanged, resulting in significantly lower computational cost compared with full retraining of the entire model. 

\begin{figure}[!thb]
\centering
\includegraphics[width=3in]{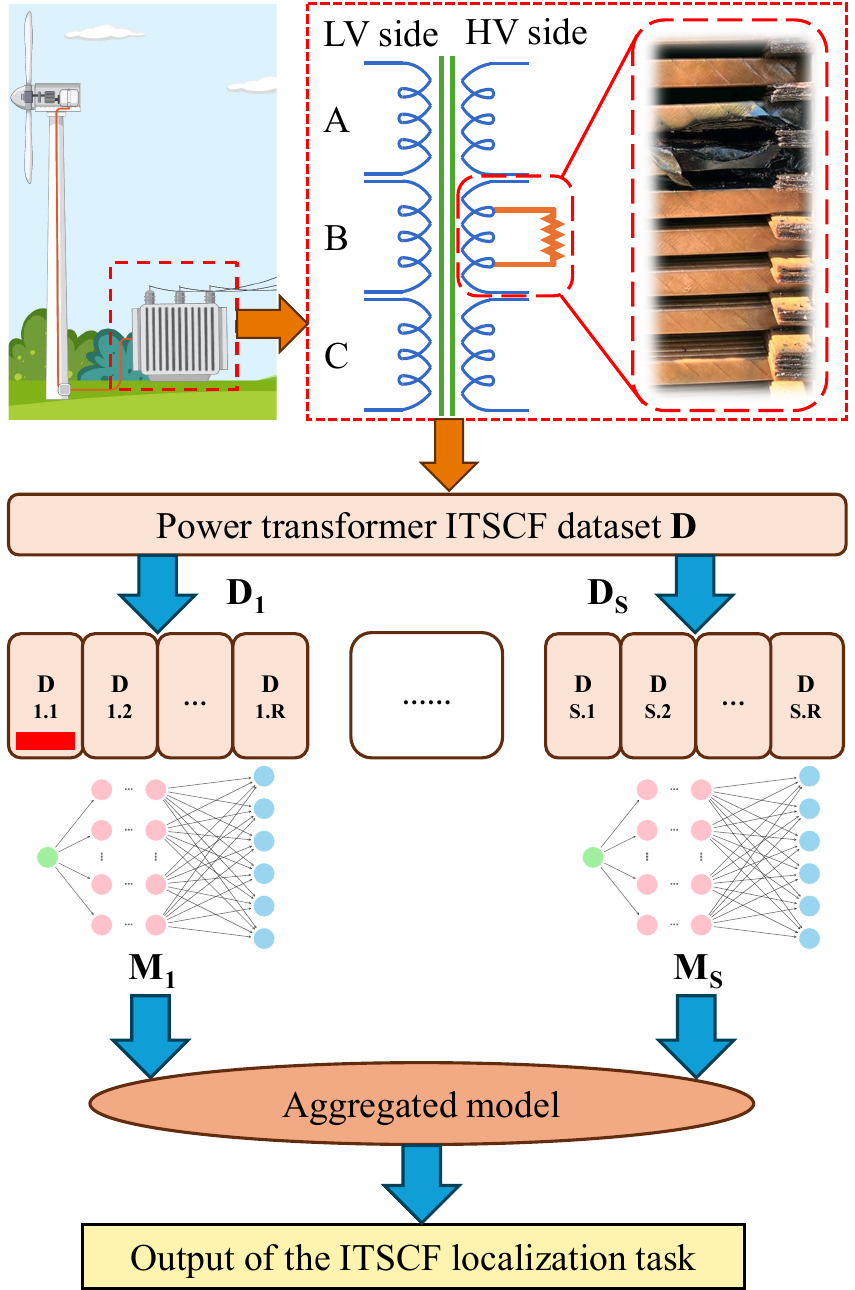}
\caption{The SISA framework for power transformer ITSCF localization.}
\label{fig:SISA_ITSCF_Framework}
\end{figure}

\subsection{SISA Framework for ITSCF Localization}
Generally, there are six types of practical MU scenarios \cite{li2025mubox}, and our power transformer ITSCF case, considering sensor failures, is classified as the scenario of depoisoning, removing the adverse effects of poisoned data. 

% In the context of the ITSCF localization task, the SISA framework enables the selective removal of EMI-poisoned current sequences that degrade the phase identification performance. 
% By retraining only the affected shards, SISA effectively eliminates the influence of harmful samples while preserving the learned information from other clean data. 
% This makes SISA a practical and efficient MU strategy for transformer condition monitoring, where sensor errors and measurement anomalies occasionally occur in electrical power systems.

As depicted in \autoref{fig:SISA_ITSCF_Framework}, the upper figure shows an ITSCF case located in phase B of the HV side, and the complete power transformer ITSCF dataset $D$ contains 48 operating conditions across six fault labels---HA, HB, HC, LA, LB, and LC---corresponding to the HV and LV sides in phase A, B or C of the power transformer connected to a wind turbine. 
The dataset is first evenly partitioned into $S$ shards ($D_1, D_2, \ldots, D_s$), each containing $48/S$ ITSCF conditions. 
Within each shard, the data are further divided into $R$ slices such that every slice includes all six fault labels, ensuring label balance and diversity. 
For instance, when $S = 2$, each shard contains 24 ITSCF conditions and is divided into $R = 4$ slices, with each slice including one sample per label. 
During the SISA training process, each shard is independently trained to obtain sub-models ($M_1, M_2, \ldots, M_s$). 
To combine the predictions from multiple shard models, a softmax-probability averaging strategy is adopted: 
Let $z_s(x)$ denote the logits of the $s$-th model for input $x$, and its softmax probability for class $c$ is
\begin{equation}
p_s(c|x) = 
\frac{\exp\big(z_s(x)_c\big)}{\sum_{k=1}^{C} \exp\big(z_s(x)_k\big)}
\label{eq:softmax_prob}
\end{equation}
The aggregated prediction probability is computed as:
\begin{equation}
\hat{p}(c|x) = \frac{1}{S} \sum_{s=1}^{S} p_s(c|x)
\label{eq:simple_avg}
\end{equation}
And the final predicted label is determined by:
\begin{equation}
\hat{y}(x) = \arg\max_{c} \hat{p}(c|x)
\label{eq:final_pred}
\end{equation}
where $S$ is the number of shard models and $C$ is the number of ITSCF localization labels.

To better reflect practical deployment, fault cases with similar severity levels were grouped into the same shard.
This represents scenarios where multiple ITSCF events may originate from the same wind farm or substation.
In such settings, sensor failures from EMI are more likely to affect samples within the same shard, making the poisoned data easier to isolate and retrain.
This setup aligns with the adaptive sharding strategy in the SISA framework, which has been shown to outperform uniform sharding in unlearning \cite{bourtoule2021machine}.

As an example, when an ITSCF condition within $D_{\text{1.1}}$ marked red in \autoref{fig:SISA_ITSCF_Framework} is found contaminated, the corresponding data shard $D_1$ is regarded as poisoned. 
Upon detection of such corrupted data, only the affected shard is retrained from scratch, while the remaining shards and their learned parameters are preserved. 
This localized unlearning mechanism enables the selective removal of poisoned data without retraining the entire model, thereby significantly reducing computational training time while maintaining the accuracy and stability of the ML task.

\section{Results Analysis} \label{sec:reaults}
\subsection{ITSCF Data Processing}
As mentioned in \autoref{sec:simulation}, the 15-second high-fidelity current measurements, sampled at 1000 Hz in the power generation mode, were divided into overlapping windows of 50 time steps to form the input sequences.
A ratio of 4:1:1 split was applied for training, validation, and testing, with the validation set used for hyperparameter tuning. 
To avoid any overlap among the three subsets, the dataset was first partitioned and subsequently shuffled within each subset.

\subsection{ML Model Structure and Hyperparameters}
All LSTM models were implemented in PyTorch 2.5.1 and trained using the Adam optimizer with a learning rate of 0.0001 for 60 epochs. 
The categorical cross-entropy loss function was employed for the classification task, and a batch size of 512 was used during training.

Each model adopts a two-layer LSTM architecture followed by a classification head.
The first LSTM layer with 96 hidden units is followed by layer normalization and dropout to improve training stability and generalization.
The second LSTM layer with 48 hidden units extracts higher-level temporal features, also followed by normalization and dropout.
Global average pooling along the temporal dimension converts the variable-length sequences into fixed-size feature vectors, which are then fed into a two-layer fully connected network: the first layer maps the 48-dimensional features to 64 dimensions using ReLU activation and dropout, and the second layer outputs logits for the six ITSCF classes corresponding to HA, HB, HC, LA, LB, and LC.
The dropout rate is set to 30\% for all dropout layers.

In this study, the ITSCF localization task is formulated as a multi-class classification task. 
To evaluate ML and MU performance, two commonly used metrics are adopted: average accuracy and computational time. 
All experiments were performed on a workstation equipped with RTX 4090.

\subsection{Experimental Results Analysis}

Since the original SISA paper has investigated the effect of slicing and demonstrated that it does not adversely affect accuracy compared with the non-sliced approach when the total number of training epochs is kept constant, while slicing offers a significant advantage in reducing retraining time \cite{bourtoule2021machine}.
Therefore, in our SISA-based ITSCF case study, the impact of slicing is not analyzed. 
Instead, we concentrate on two key issues: (1) how the number of shards influences the aggregated model accuracy and the computational efficiency of SISA unlearning, and (2) how single or multiple ITSCF conditions, when poisoned by sensor failures, affect model accuracy before and after full retraining or SISA-based unlearning.
To meet the two objectives, we assume that one of the detected poisoned ITSCF conditions is confined to the first slice of its respective shards. 
This setting eliminates the influence of slicing and focuses solely on the two mentioned issues.

For the first problem, to compare the accuracy and training time with and without poisoned data across different shards, EMI-induced sensor failures were introduced into one ITSCF condition on the LA side.

\begin{figure}[!thb]
\centering
\subfloat[Accuracy vs. number of shards]{%
    \includegraphics[width=2.9in]{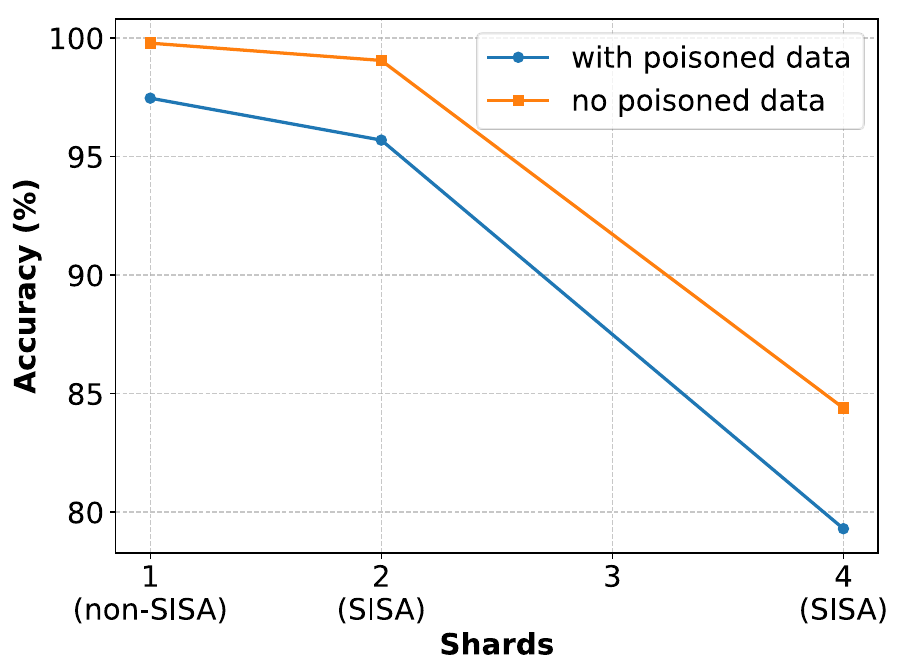}
    \label{fig:Acc_SISA}
}
\hspace{0.1in}
\subfloat[Training time vs. number of shards]{%
    \includegraphics[width=2.9in]{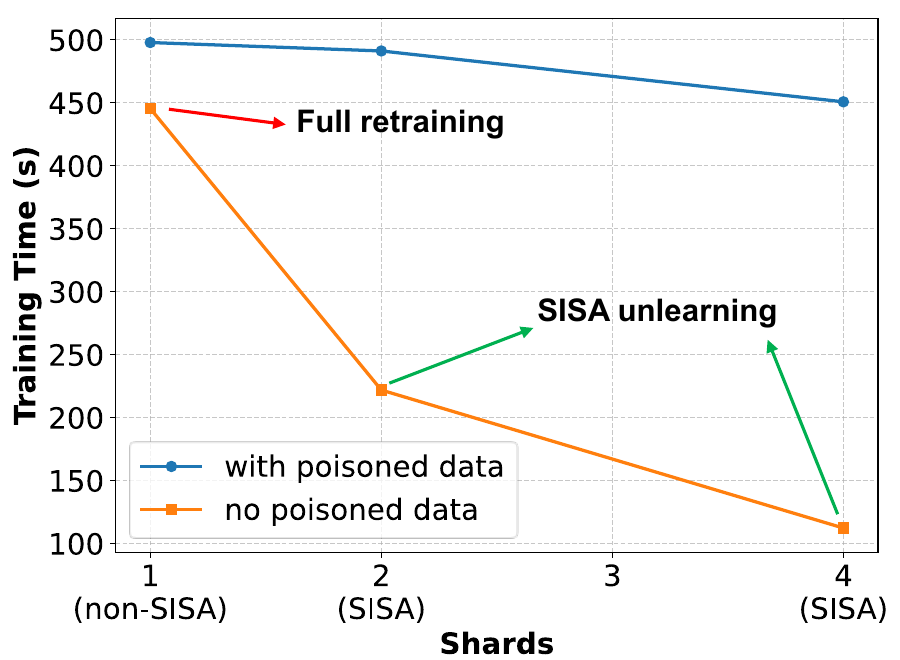}
    \label{fig:Time_SISA}
}

\caption{Training accuracy and time on different numbers of shards, i.e., $S$. 
If $S=1$, it is a non-SISA model, and the training process removing poisoned data is full retraining.}
\label{fig:Acc_Time_SISA}
\end{figure}

As shown in \autoref{fig:Acc_Time_SISA}(a), removing poisoned data uniformly improves model accuracy for all configurations. 
In the non-SISA model with $S=1$, the accuracy training with poisoned data is 97.46\%, and it increases to 99.78\% when full retraining with clean data.
Compared with the non-SISA model with a single shard, the SISA model with two shards achieves slightly lower accuracy, from 95.69\% to 99.05\% when eliminating poisoned data, which is consistent with the original SISA paper indicating that sharding into a few shards causes no remarkable degradation in accuracy. 
However, when the number of shards is set to four, the accuracy drops more sharply to 79.30\% and 84.39\% with or without poisoned data.
This decline occurs because the training dataset contains a limited number of ITSCF conditions, and dividing the dataset into four or more shards could reduce the diversity of training data within each shard, and increase the likelihood of misclassifying similar current patterns between HV and LV sides.

The training time results in \autoref{fig:Acc_Time_SISA}(b) further highlight the computational advantage of the SISA-based MU framework. 
In the non-SISA configuration, full retraining after removing poisoned data slightly reduces the training time from 497.8~s to 445.4~s due to a smaller dataset.
By contrast, if the whole dataset is partitioned into two or four shards, SISA unlearning requires retraining only on the affected shard instead of the entire model.
When the number of shards is set to two, the SISA unlearning process reduces the retraining time to 221.8~s, and additionally decreases to 112.2~s when four shards are applied, achieving speed-ups of  2.01$\times$ and 3.97$\times$, respectively.
Although the training time is short due to the limited data availability, the proposed design still achieves a notable reduction in retraining time, demonstrating that the SISA-based MU framework maintains satisfactory diagnostic accuracy while significantly improving computational efficiency for power transformer ITSCF localization. 
With larger and more complex realistic datasets, the improvement in computational efficiency is expected to become more pronounced and significant.

% \JZ{since the computational time is low in all cases, it is better to clarify that this is only a toy case. With larger amount of data, the time reduction will be more significant.}

% \begin{figure}[!thb]
% \centering
% \includegraphics[width=2.7in]{figures/Acc_Time_SISA.pdf}
% \caption{Accuracy and training time on different number of shards, i.e., $S$. 
% If $S=1$, it is a non-SISA model, and the training process removing poisoned data is full retraining. \JZ{Add the subfigure caption.}}
% \label{fig:Acc_Time_SISA}
% \end{figure}

\begin{figure}[!thb]
\centering
\includegraphics[width=3.6in]{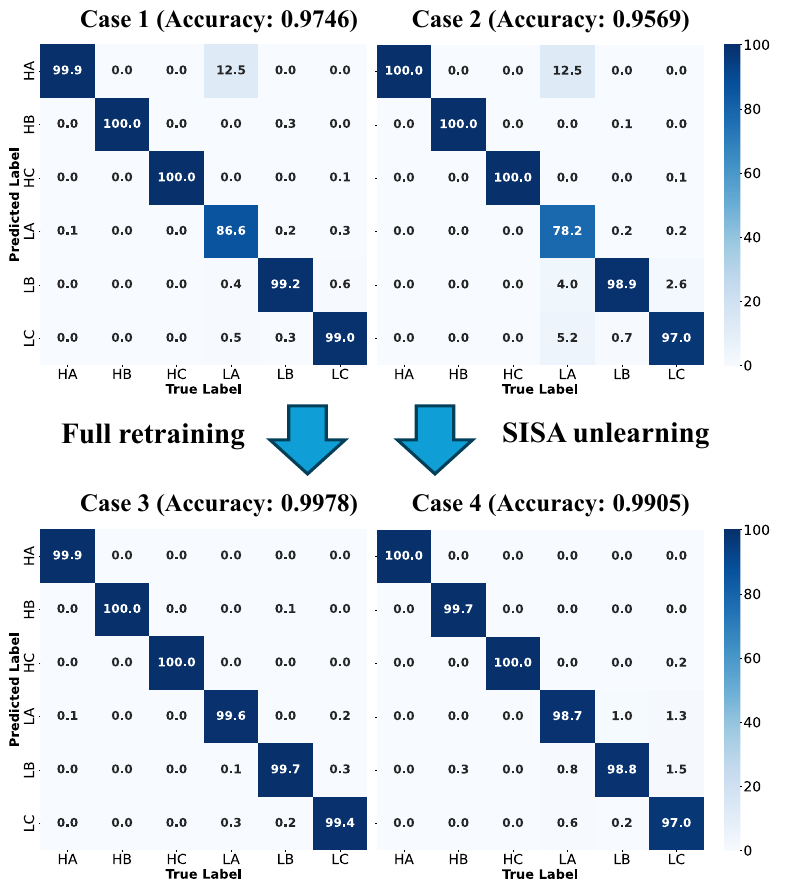}
\caption{Comparison of confusion matrices for four cases under 1 ITSCF condition with poisoned data caused by sensor failures.}
\label{fig:CM_SISA_F1}
\end{figure}

\begin{figure}[!thb]
\centering
\includegraphics[width=3.6in]{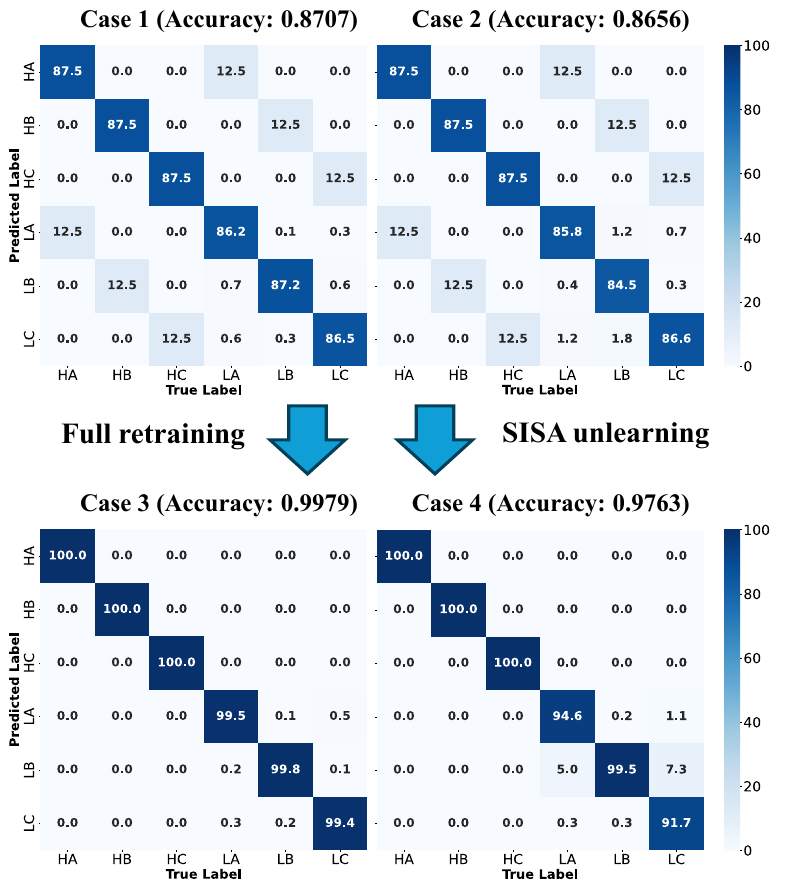}
\caption{Comparison of confusion matrices for four cases under 6 ITSCF conditions with poisoned data caused by sensor failures.}
\label{fig:CM_SISA_F6}
\end{figure}

To explore the second issue, we adopt $S=2$ since it achieves a reasonable compromise between model accuracy and training time; meanwhile, we evaluate four cases:
\begin{enumerate}[label=\arabic*.]
    \item Case 1: Non-SISA training with poisoned data.
    \item Case 2: SISA training with poisoned data.
    \item Case 3: Non-SISA full retraining without poisoned data.
    \item Case 4: SISA unlearning without poisoned data.
\end{enumerate}

As illustrated in \autoref{fig:CM_SISA_F1} and \autoref{fig:CM_SISA_F6}, two representative configurations of poisoned data, with one or six ITSCF conditions poisoned, are compared through confusion matrices. 
The results clearly demonstrate that both full retraining and SISA unlearning effectively restore the classification accuracy degraded by sensor failures. 

In \autoref{fig:CM_SISA_F1}, the sensor failure is introduced into one ITSCF condition on the LA side, which causes a noticeable drop in the recognition accuracy of LA.
For the training with poisoned data in both Case~1 and Case~2, the LA label is misclassified as HA with a rate of 12.5\%, because the current waveforms of ITSCF conditions on HA and LA exhibit comparable characteristics.
After the poisoned data are removed, the accuracy in Case~3 and Case~4 rises to nearly 100\%, confirming that both full retraining and SISA unlearning can recover the diagnostic performance.

In \autoref{fig:CM_SISA_F6}, one ITSCF condition from all three phases on both HV and LV sides is contaminated by sensor failures, causing a total of six poisoned conditions.
This configuration results in decreased classification accuracies across all six labels, since the ITSCF current signals from the HV or LV side in the same phase are quite similar, making them prone to be wrongly classified.
Once the contaminated samples are removed, the average accuracy significantly improves from around 87\% to more than 97\%.

Through all confusion matrices, especially the Case~2 in \autoref{fig:CM_SISA_F1} and the Case~4 in \autoref{fig:CM_SISA_F6}, it is also observed that the three LV side phases exhibit more frequent misclassifications than the HV side, revealing the strong similarity among the three LV side phases (LA, LB, and LC).
Furthermore, the accuracy in Case~2 and Case~4 within the SISA framework is always lower than the corresponding non-SISA model of Case~1 and Case~3, which is consistent with the original SISA study that reported modest accuracy degradation due to sharding. 

Overall, these results verify that the proposed SISA-based MU framework can effectively eliminate the negative influence of poisoned data caused by sensor failures, while achieving comparable accuracy to full retraining at considerably lower computational cost.

\section{Conclusion} \label{sec:conclusion}
This paper presented a SISA-based MU framework for the power transformer ITSCF localization task considering sensor failures. 
By partitioning the whole dataset into multiple shards and retraining only the affected shard, the proposed method efficiently removes the impact of poisoned data while avoiding full retraining.
Experimental results demonstrate that SISA unlearning achieves almost the same accuracy as full retraining, improving accuracy to more than 97\% after removing poisoned data, while substantially reducing retraining time by up to 3.97$\times$ with four shards.
The results also reveal that excessive sharding may reduce accuracy due to limited data diversity, and misclassifications mainly occur among the same phase on each side and three phases on the LV side.
Overall, the proposed framework provides a practical, computationally efficient unlearning strategy for power transformer ITSCF localization. 
In future work, other MU methods could be explored in similar data-driven condition monitoring systems.

% \JZ{please cite our own ITSCF papers, including NAPS paper, Jingyi's conference and journal papers.}
% \JZ{The results are all expected. It is not very interesting. I think we should focus more on machine unlearning algorithm improvement in the future. Because we don't have large datasets to train. But new algorithms can always be evalutated with toy examples.}

\bibliographystyle{IEEEtran}
\bibliography{refs}

\vfill
\end{document}